\documentclass[letterpaper]{article} 
\usepackage{aaai2026}  
\usepackage{times}  
\usepackage{helvet}  
\usepackage{courier}  
\usepackage[hyphens]{url}  
\usepackage{graphicx} 
\usepackage{natbib}  
\usepackage{caption} 
\usepackage{algorithm}
\usepackage{algorithmic}

\usepackage{booktabs} 
\usepackage{newfloat}
\usepackage{listings}
\usepackage{tabularx}
\usepackage{colortbl}
\usepackage{colortbl}
\usepackage{multirow}
\usepackage{makecell}
\usepackage{graphicx}
\usepackage{longtable}
\usepackage{subcaption}
\usepackage[table]{xcolor}
\usepackage{colortbl}
\usepackage{xcolor}
\usepackage{fancyvrb}
\usepackage[table]{xcolor}
\usepackage{colortbl}
\usepackage{xcolor}
\usepackage{fancyvrb}
\definecolor{rowhighlight}{RGB}{240,248,255} 

\definecolor{codegreen}{rgb}{0,0.6,0} 
\definecolor{codegray}{rgb}{0.5,0.5,0.5} 
\definecolor{codepurple}{rgb}{0.58,0,0.82} 
\definecolor{backcolour}{rgb}{1,1,1} 
\definecolor{blackfont}{rgb}{0,0,0} 

\colorlet{rowhighlight}{codegray!12}
\newcommand{\highlightrow}{\rowcolor{rowhighlight}}
\DeclareCaptionStyle{ruled}{labelfont=normalfont,labelsep=colon,strut=off} 
\floatstyle{ruled}
\newfloat{listing}{tb}{lst}{}
\floatname{listing}{Listing}
\title{Synthetic Sources?: Auditing Generative Search Engine Citations for Evidence of
AI-Generated Sources}
\author{
    Mowafak Allaham,
    Nicholas Diakopoulos
}
\affiliations{
    Northwestern University\\

}

\usepackage{bibentry}

\begin{document}

\maketitle

\begin{abstract}
The growing accessibility of Large Language Models via conversational interfaces capable of responding to users' questions by drawing on, synthesizing, and citing information from the web (i.e., Generative Search Engines) has simplified the information-seeking process for users. 
However, with the proliferation of AI-generated content on the web, it is unclear whether these engines can reliably omit citing synthetic sources (i.e., AI-generated sources). Should these engines be unable to do so, this puts users at risk of harm by treating information from AI-generated sources synthesized in responses of generative search engines as equivalent to information from authoritative or official sources. 
In a step towards identifying whether AI-generated sources are being cited by these engines, this work presents an audit of four generative search engines (ChatGPT, Copilot, Gemini, Perplexity) using a total of 712 real-world human-generated queries spanning domains of public importance: politics, health, and the environment. Our findings show evidence of AI-generated sources being cited across all four generative search engines ($\sim$16\% of cited sources) and identifies key source web domains these sources belong to that are frequently cited across these engines and topics. In addition, we observed that generative search engines include a somewhat narrow set of repeatedly cited domains while predominantly surfacing a large number of minimally cited domains in responses to users' queries. These findings contribute to the growing body of work on assessing the risks of generative search engines with the objective of increasing public awareness of their limitations and encouraging appropriate measures to improve information quality and governance of these systems.
\end{abstract}


\section{Introduction}

People turn to search engines (e.g., Google or Bing) to gather and evaluate information, form opinions, and make decisions on a wide range of topics they find important on the personal and societal levels \cite{rieger2024nudges,dutton2017search,ghosh2018searching,moraes2018contrasting,marchionini2006exploratory}. 
However, the growing accessibility of search-enabled large language models (LLMs) via conversational interfaces capable of responding to users' queries by drawing on and synthesizing information from the \textit{web} is transforming how people seek information online. In this work we refer to these tools, which include popular search-enabled LLMs such as ChatGPT or Gemini, as \textit{generative search engines} (GSE). Generative search engines simplify the search process for users by shifting their attention away from clicking and visiting multiple links on a traditional search results page toward consuming a single synthesized response substantiated with in-text citation to various sources -- 
a source in this context refers to a unique web URL that represents an item online (e.g., blog post, an article, image, or video) that is being cited by generative search engines.


Experts and journalists have expressed growing concerns about the potential of generative search engines in relying more on AI-generated sources as a means to compensate for gaps in access to authoritative sources of information and to fill information vacuums (i.e., ``data voids'') when responding to users' queries \cite{russell2025ai-620,alyukov2025llms,tripodi2025data,bansal2025moderateAIOverviews}. While the epistemic quality of AI-generated content could be a contentious point of debate, as not all AI-generated content is necessarily low quality \cite{nishal2024envisioning-891}, the AI-tools used for generating such content (e.g., large language models) have their biases \cite{zhang2025systematic,gallegos2024bias} and are susceptible to generating hallucinations and inaccurate information \cite{bommasani2021opportunities,deng2025cram,luo2025unsafe}. This, in turn, may affect the information quality of the output produced by these tools. In addition, the sourcing mechanisms of generative search engines for information online and the choices these engines make at the system and information levels as part of generating the final answer, such as when to include sources in responses, what sources to include or exclude, and the quality and credibility of these sources, illustrate the opaqueness of these engines, which creates additional opportunities for AI-generated sources that may include false or misleading information to compete with accurate ones \cite{guardian2026googleAI,roy2025ai,tagesanzeiger_ki_fake_news_2025,newsguard_ai_tracking_center}. This ambiguity in information sourcing mechanism exposes responses of generative search engines to potential manipulation for personal or corporate gains (e.g., brands promoting AI-generated content onlie to rank themselves in first place in ``best X'' lists) \cite{allsopp2025bestlists} or get exploited by bad actors, through knowledge injection or poisoning attacks \cite{soudani2025enhancing,an2025rag,deng2025cram}, to amplify sources with low-quality AI-generated content in the pool of sources on which generative search engines rely on, potentially degrading the quality of generated responses. 

Therefore, if the source content cited and used by generative search engines as part of the response generation process is AI-generated, given the aforementioned issues associated with this type of content, users may be at risk of unknowingly consuming low-quality information. This is particularly consequential for topics of individual (e.g. health) or societal (e.g. politics, environment) importance, which could misinform beliefs and actions or influence public opinion and upset political processes \cite{li2024generative,van2025does}.

To better characterize the nature and extent of this problem, it is imperative as a first step to identify whether or not generative search engines cite sources with AI-generated content. To this end, this research undertakes an audit of the sources cited by four popular generative search engines, which include the search-enabled Chatbot interfaces of ChatGPT, Copilot, Gemini, and Perplexity \cite{ebu2025newsintegrity}. We first identify the range of source web domains (i.e., web domains that sources belong to) these engines rely on to respond to users' queries, tabulate their frequencies, whether or not the source content corresponding to cited sources from these web domains is AI-generated, and then examine the prominence of such content among cited sources. To this end, we pursue the following two research questions: (1) \textbf{what are the prevalent source web domains generative search engines cite sources from when responding to users' queries?} and (2) \textbf{how prevalent is AI-generated sources among those cited sources?} We address these questions by auditing answer engines through a socio-technical lens \cite{solaiman2023evaluating,metaxa2021auditing,allaham2025informing} that extends beyond assessing the technical artifacts of AI systems to reflect real-world users' queries spanning three topics of public importance: politics, health, and the environment \cite{li2024generative,van2025does}


Using a closely curated sample of 712 English language queries focusing on general knowledge questions spanning the topics of politics, health, and the environment, we audit the cited sources of the aforementioned four generative search engines for evidence of AI-generated sources by first submitting each query in our sample through the user interface (UI) of each engine, then (if present) scraping the cited sources and their respective content in each generated response, and finally using an AI-detection tool to identify whether the source content in these sources is AI-generated or not. 

To assess whether source content from multiple cited sources is AI-generated, we use Pangram \cite{emi2024technical}, an AI-generated-text detection tool, and observe a considerable share of source contents of cited sources (approximately 16\% of unique cited sources across all providers) on GSE to be AI-generated, with responses to environment-related queries exhibiting the highest proportion of cited source content that is AI-generated. This is a significant finding in the context of auditing information on such engines, as the AI-generated source content being cited is prone to be of lower quality, especially given that more reputable news websites were reported to disallow AI crawlers than ones containing misinformation \cite{steinacker2025misinformation}, which may disproportionately (or possibly negatively) influence the content produced by generative search engines that is presented to users in responses.

Our findings contribute to the growing body of research investigating information-related risks posed by generative search engines to users and society at large \cite{luo2025unsafe} by presenting evidence of AI-generated sources being cited by four popular generative search engines \cite{ebu2025newsintegrity}, highlighting limitations in the reliability of these engines especially when responding to queries on topics of public importance (politics, health, and the environment). In addition, by mapping and tabulating the frequencies of source web domains cited by these engines, our findings reveal the disproportionate sourcing practices of these engines along with the subset of source web domains they frequently rely on to substantiate their answers. These findings may inform practitioners, including developers and risk-assessors, about the prevalence of AI-generated sources, both across and within individual generative search engines, and the source web domains from which they originate. Such insights could inform governance efforts 
looking to improve generative search engines by identifying (and potentially omitting) sources with AI-generated content in citations, particularly if found to be of lower quality. Ultimately our findings may help take steps towards enhancing the overall quality of information used by these systems to generate responses.
\section{Related work}

Generative search engines are increasingly becoming the primary systems for information seeking, potentially surpassing traditional search engines \cite{liu2023evaluating,chapekis2025google}. However, these systems largely remain opaque in how they source, synthesize, and present information from the web. This lack of transparency, coupled with the rapid deployment of these systems, raises concerns about their potential impacts on individuals and society, and particularly on topics of critical importance such as health and politics \cite{li2024generative}. One established strategy for interrogating the inner-workings of AI and algorithmic systems, such as generative search engines, and understanding them from the outside in is: algorithmic audits \cite{metaxa2021auditing}.

The ambiguity in how generative search engines operate and sources content from the web have been obscuring information quality issues, which have tremendous negative consequence on users, such as experiencing difficulty in assessing the reliability of generated responses \cite{venkit2024search,li2024generative,yang2025news}. As a result, audits have previously evaluated the quality of citations and responses of generative engines along numerous dimensions, including information verifiability \cite{liu2023evaluating}, response factuality \cite{venkit2024search}, source attribution (and mis-attribution) \cite{strauss2025attribution,kelly2023bing,urman2025silence}, and political bias \cite{dai2025media,minici2025auditing,rozado2022_political_orientation_chatgpt}. Empirical findings from these audits consistently reveal clear limitations in the reliability and safety of these systems \cite{li2024generative,kuai2025ai,luo2025unsafe,roy2025ai}. For instance, studies report that generative answer engines often respond to questions they cannot accurately answer rather than declining to respond, resulting in incorrect or speculative outputs, fabricated links, inaccurate citations, and the inclusion of unsupported or misleading claims \cite{jazwinska_chandrasekar_2025,liu2023evaluating}, which at times could put users at risk of harm as reported on in the media in the case of Google AI Overviews putting users at risk of harm by sharing misleading health advice with inaccurate information from AI-generated summaries \cite{guardian2026googleAI}. More problematically, other studies have highlighted how generative search engines generate citations that are not clickable and consistently present information to users without conducting live web search, thus relying on the LLM's internal knowledge for answers \cite{strauss2025attribution,liu2023evaluating}. This in turn has its negative implications on users being exposed to false or misleading information, as LLMs are prone to hallucination \cite{ji2023survey}. 


Beyond these risks, studies have also examined the societal implications of these systems \cite{memon2024search,bender2021dangers}, including their potential to influence public opinion \cite{newsguard_ai_tracking_center,dai2025media}, create ``Chat-Chamber'' effects \cite{jacob2025chat}, engage in forms of information homogenization or gate keeping \cite{lindemann2025chatbots,brantner2025sourcing,vranic2025global}, and exposing users to information that are ``less diverse, more biased, and unverifiable—yet expressed in a confident manner'' \cite{memon2024search}. Collectively, these findings characterize the risks generative search engines may have on individuals and society as a result of the content and sources they feature in their responses. One emerging type of content that may further exacerbate the negative impacts of these systems on users and society is the growing proliferation of AI-generated content on the web \cite{russell2025ai,shaib2025measuring}.

Recent work has begun to develop qualitative frameworks and computational tools for characterizing such content more precisely \cite{emi2024technical,shaib2025measuring}. For example, \citet{shaib2025measuring}, introduced a taxonomy of ``AI slop'', defined as low-quality AI-generated text, to measure its prevalence, highlighting that existing qualitative judgments of AI content lack consistent operationalization \cite{shaib2025measuring}. Another example that leverages computational approaches is Pangram\cite{emi2024technical}, a transformer-based classifier that has been trained on extensive dataset of human and synthetic texts and shown to accurately discern AI-generated from human-authored content \cite{emi2024technical}.

Overall, we find that the existing literature on auditing generative search engines has not yet explored the evaluation of cited sources for the presence of AI-generated content in responses to users queries across topics of public importance, such as politics, health, and the environment. This gap presents an opportunity to leverage emerging computational methods, like Pangram, to audit the quality of sources cited by answer engines as described in subsequent sections. 

\section{Data}

\subsection{Curation \& Selection for Queries on Politics and Health}
To curate a dataset that reflects the types of queries users would submit to generative search engines specifically (rather than queries they might submit to a traditional search engine), we start by leveraging queries in the English language from the Search Arena dataset \cite{miroyan2025search}.  This dataset contains 24,069 queries and conversational interactions with generative search engines, collected between March 18, 2025 and May 8, 2025, across 70 languages from 11,650 users across 136 countries, capturing a broader range of user search intents and topics. To conduct an in-depth analysis of the queries, the authors of the Search Arena dataset used a mixed annotation approach that involves human and GPT-4.1. First, they used 100 queries as a seed to tune the annotation process for GPT-4.1. Next, they validated the agreement between human and GPT-4.1 annotations on a sample of 150 multilingual queries, achieving  a Cohen's kappa of 0.812 which indicates a strong agreement between model- and human-annotated labels \cite{miroyan2025search}. This annotation process groups queries in the dataset into nine intent categories: Factual Lookup, Information Synthesis, Analysis, Recommendation, Explanation, Creative Generation, Guidance, Text Processing, and an Other category (see Appendix \ref{q-intent} for detailed description of these categories). Seven out of the nine categories are informational in nature constituting 87.4\% of the dataset. Our focus in this work is on informational intents and so we exclude the Creative Generation and Other categories as one captures queries requesting original creative outputs, including fictional scenarios and poetry, and the other represents queries that are malformed, declarative in nature, or lacking in meaningful user intent.

Because the dataset includes both single and multi-turn interactions between users and generative search engines, we chose to unify the structure of the data by transforming multi-turn interactions into single-query instances, as some multi-turn interactions are multilingual (i.e., they involve queries in English alongside queries in other languages over the course of an interaction). Accordingly, for each interaction session between a user and the system (i.e., the ``assistant"), we retained only ``user" messages that are in English. For the remainder of the analysis, we treat each query as an independent unit of input to the system, rather than considering the sequence in which queries were submitted, even in cases where multi-turn interactions were conducted entirely in English. This choice enables the retrieval and analysis of the sources generative search engines cite in responses to individual queries, rather than sources cited in multi-turn conservational interactions between users and the system (a limitation of our study that we highlight the implications of in the limitations section). After de-duplicating queries, this transformation process resulted in a total of 11,185 single-turn and unique queries in English. Next, to determine which of these queries are on politics, health, or the environment, we first had to determine the topical scope of these queries and then assess their relevance to our topics of interest. 

Using BERTopic \cite{grootendorst2022bertopic}, a topic modeling technique that leverages transformer-based embeddings to produce interpretable topics, we followed the open-source pipeline developed by the authors of Search Arena for identifying common topics from queries in the dataset \cite{tang2025arenaexplorer}. Our decision to use BERTopic stems from its approach, which enables the assignment of queries to topics through a zero-shot method while still allowing qualitative evaluation of the generated topics. In this pipeline, queries are first transformed to high-dimensional vector embeddings, which is then used to identify topic clusters.
This process resulted in queries being grouped into 76 topics spanning a wide range of domains, including sports, music, gaming, financial advice, legal advice, and others. To map these topics, and the queries associated with them, back to the three topic categories of interest (politics, health, and the environment) we operationalize the definition for each topic category based on prior research describing these topics in the context of online search. Queries on politics involve seeking, monitoring, or evaluating information about political actors, parties, issues, or events. Such queries are typically motivated by informational needs including uncertainty reduction, belief validation, or preparation for political decision-making \cite{menchen2023searching,dutton2017search}. Health-related queries are characterized by seeking information, clarification, or guidance concerning health issues, symptoms, diseases, and medications, as well as wellness practices and recommendations \cite{jia2021online,zhu2025cancer,swire2020public}. As for queries on the environment, they typically include questions aimed at understanding climate change, its causes, consequences, and potential mitigation or reversal measures, as well as exploring earth systems and alternative energy sources \cite{de2024climateq}. 

Using the definitions described above, authors conducted a manual review of a random sample of five representative queries from each of the 76 topics produced by the topic modeling pipeline. This review enabled us to identify eight topics (two on health and six on politics) comprising 1,325 queries that fall within the domains of politics and health. However, no queries were found to be related to the environment other than weather inquires (e.g., ``what is weather like today?'') which are not considered relevant under our definition of environment-related topics. Because the dataset includes queries from users across 136 countries, authors have subsequently discussed and then performed a joint open review of these queries to ensure relevance to U.S. contexts. We chose to exclude queries that are not relevant to the U.S. because (1) all four generative search engines under audit are developed primarily in the United States, and (2) the evaluation of the quality of citations included in responses to queries on topics that are relevant to other regions requires domain and contextual expertise specific to the information ecosystem in these regions, which the authors do not claim to have the expertise in. It is worth noting that this process was not uniformly trivial across all queries, as it required reading each query and attempting to identify keywords that might indicate irrelevance. For instance, encountering terms such as ``Tesco'' (a British multinational grocery store) or ``Zapatista Army of National Liberation (EZLN)'' (a political group in Mexico) required additional information lookup by the authors to determine whether the query was relevant. While this presents a limitation to our approach in excluding queries, our experience may help inform future researchers to anticipate the presence of such instances when dealing with datasets that are multilingual or have been collected from users across multiple nations. For instance, we excluded queries that primarily focus on regions beyond the U.S. (e.g., ``How does supply chain bifurcation impact  Australia''; ``why only germany not embrace Modern Monetary Theory in rich nation?''), political queries discussing issues that are not directly relevant to the U.S. (e.g., ``Why is Taiwan a country?''; ``Did the premier of China Xi encourage the youth to forego college and instead go for manufacturing jobs?''), or health related queries or topics specific to other countries (e.g., ``Talk me through the next procedure may see in UK for pediatric patient"). As a result, 432 queries ($\sim$32\% of 1,325 queries) were retained that are on politics (N=257) and health (N=175) that fit our query selection criteria.

Queries on health in our sample include one topic cluster focusing on seeking information about supplements (e.g., ``Which foods Contain all essential amino acids? Protein Drinks?''), vaccines (e.g., ``should children get their second mmr vaccine early among a measles outbreak?''), and medications (e.g., ``what do you need to do to make the effects of bromantane to last''), including questions about drug-drug interactions (e.g., ``can i use zofran and bisacodyl together''), and the other topic cluster includes informational queries related to mental, physical, or physiological health concerns (e.g., ``what are the effects of nose breathing and mouth taping on face/jaw structure?''; ``How to prevent nervous tongue scratching against the teeth?''). As for queries on politics, the topics spanned over U.S. political leadership and governing institutions, with a focus on President Donald Trump and his administration (e.g., ``Do you think Donald Trump and his administration members participated in insider trading?''), evaluating the consequences of a public policy (e.g., ``Which universities and states will be most impacted by the denial of funds on science?''), U.S. imposed tariffs and trade wars (e.g., ``What is the effect of the ongoing tariffs enacted by trump''), informational queries about political actors aimed at uncertainty reduction (e.g., ``Who is the current U.S. ambassador to Israel, and how might they engage in diplomacy?"; ``When was the last time a sitting president visited Waco, TX?''), queries pertaining to belief formation (e.g., ``Why the young men shifted towards the right wing''; ``Fascist authors and philosophers that opposed electoralism''; 
``Authors who claimed that technology could usher in a communist economy''), and other inquiries concerning historical political events and outcomes (e.g., ``Who was 1st president in us'').

\subsection{Curation \& Selection for Queries on the Environment}
Due to the absence of environment-related queries in the Search Arena dataset we rely on queries from the Climate Q\&A dataset \cite{de2024climateq} for this topic. This dataset includes 3,425 single-turn queries, collected between April and March of 2023, distributed across 16 languages submitted by users mostly from France that are interacting with an answer engine (i.e, a Chatbot) that responds to environment-related queries regarding users' actions on nature (e.g., transportation or food choices) using over 14,000 pages of scientific literature sourced from the Intergovernmental Panel on Climate Change (IPCC) and Intergovernmental Science-Policy Platform on Biodiversity and Ecosystem Services (IPBES) reports\cite{delacalzada2024climateqabridginggapclimate}. Queries in this dataset are distributed over fifteen topic categories (see Appendix \ref{c-q-n-a}) ranging from questions about earth systems (e.g., ``What will be the sea level rise in 10 years'') to inquiries about the potential actions companies could take to mitigate climate change (e.g., ``what need to be done inside companies to fight climate change ?'').

Unlike the Search Arena dataset where the language is annotated and provided by the dataset creators, in this dataset we employed Python's \textit{langdetect} library on all queries and excluded the ones that are not in English. As a result, we retain 77.9\% of all queries (N=2,668) as the sample of queries in English. Next, we sampled 20\% of the queries in English (N=534) for further analysis. All queries were reviewed for regional relevance, following the same procedure applied to the Search Arena dataset. Accordingly, queries explicitly focusing on environment-related impacts, policies, or mitigation strategies for countries or regions other than the U.S. were excluded (e.g., ``what are the top 3 cost effective solutions for mitigating climate change in France?''; ``What will be tje sea level in france in 2100?''; ``what will be climate impacts on switzerland?''). Based on this screening criteria, we retained 52.4\% of the queries (N=280) that also preserve the relative distribution of topic categories present in the original sample, as shown in Appendix \ref{c-q-n-a-sample}.

By aggregating queries from Search Arena and Climate Q\&A datasets, our final sample (N=712) encompass queries on politics (N=175; 24.5\%), health (N=257;36.1\%), and the environment (N=280; 39.3\%). In the next section we describe how they are used to audit four generative search engines (ChatGPT, Copilot, Gemini, and Perplexity).
\section{Methodology}
\subsection{Scraping citations from generative search engines}

Each query in the curated dataset (described in the previous section) was submitted to four generative search engines: ChatGPT (with web search enabled), Copilot, Gemini, and Perplexity using the user interface for each system. While these systems each have their own APIs that would make information collection simpler, prior studies have highlighted differences between responses to API requests and those submitted via web interfaces \cite{schatto2025chatgpt}. We opted to capture the cited sources in the output these engine display to users in response to their queries as this is a better reflection of what actual users would see and how they could be impacted. To this end, we developed a custom web scraper in Python that uses Playwright \cite{playwright2026}, a framework for web testing and automation, to simulate user behavior on generative search engine. 

For each generative search engine, the scraper navigates to the website corresponding to each engine, inserts the query in the text area of the search bar, activate the web search (if needed as is the case for ChatGPT), and then monitors changes in the HTML source code of the web page while the answer is being generated until no more changes are observed in the source code. Once an answer is fully generated, the scraper then clicks on the button corresponding to the sources (or citations) to view the full list of cited sources in the response. Although some engines display \textit{related} sources beyond the cited sources, we only focus on scraping the sources cited by the generative search engine in the response. Accordingly, for each cited source in the response, we scrape the source title, source URL, and source snippet (i.e., short description of the source, if available).

The final dataset collected includes 26,266 unique URLs (i.e., sources) linking to 7,675 unique web domains featuring online content that were cited by the four generative search engines. 
In the following section, we describe the procedure used to audit generative search engine citations for evidence of AI-generated content.

\subsection{Detecting AI-generated content}
The first step to identify AI-generated content among the 26,266 sources cited by generative search engines is to scrape the content for each source. To this end, we developed a Python script that programmatically visits each of the 26,266 source URLs and uses newspaper3k \cite{newspaper3k_docs} to extract the main body text from each webpage. Out of the total URLs in our dataset, we successfully scraped the content of 19,154 (72.9\%) URLs. The remaining URLs could not be scraped because they linked to other data formats beyond text, including PDFs and images (which we report on in the Limitations section). In addition, other URLs were not scraped because the content was either inaccessible or had been removed, per the HTML response code.

Next, we assess the performance of two prominent AI-detection tools, Pangram \cite{emi2024technical} and GPTZero \cite{adam2026gptzero}, in detecting AI-generated content. Pangram is built on top of a transformer-based classifier trained on a large corpus of human and AI-generated text, which has been shown to achieve substantially lower false positive and false negative rates than prior detection approaches \cite{russell2025ai-620}. In comparison, GPTZero uses a hierarchical classifier architecture that offers a ``state-of-the-art industrial AI detection solution, offering reliable discernment between human and LLM-generated text'' \cite{adam2026gptzero}. Despite the overlapping objective for both tools in detecting AI-generated content in text, they differ in their reporting of prediction scores. Pangram classifies text based on its likelihood of being either Highly Likely AI (i.e., AI-generated), Unlikely AI (i.e., human), or Mixed (i.e., AI-assisted text that is \textit{possibly} or \textit{likely} to involve the use of AI) with confidence scores ranging from 0 (least confidence) to 1 (most confident). Whereas GPTZero assigns percentages that add up to 100 that represent the tool's confidence in the likelihood in the classification of an entire text across all three categories: AI, Human, or AI-Human Mix. 

Based on the objective of this study in detecting AI-generated content in cited sources by generative search engines, it is critical to formally assess the performance of these AI-detection tools based on their false positive (i.e., human-generated content being classified as AI-generated) and false negative (i.e., AI-generated content being classified as human generated) rates. We decided to conduct this assessment to ensure that we select the AI-detection tool that has the lowest false positive rate in order to avoid mis-classifying human-generated content in our corpus as AI-generated, which consequently would inflate our results. At the same time, and depending which AI-tool we proceed with, it is important to identify the false negative rate so the results are interpreted as a lower bound to the prevalence of AI-generated content cited by generative search engines. It is important to recognize that, given our conservative selection criteria for the AI-detection tool, the reported lower bound in the subsequent section of AI-generated content remains contingent on the specific tool employed, as discussed further in the Limitations section.

In our evaluation of both Pangram and GPTZero on a curated dataset of 200 human-authored texts from the RAID benchmark \cite{dugan2024raid} and 200 curated AI-generated texts (as further described in \ref{evals} and illustrated in Figure \ref{prompt-text-generation} of the Appendix) spanning the categories of research abstracts, wikipedia pages, reddit posts, and news articles, we find that Pangram and GPTZero have 100\% accuracy in correctly detecting human-authored and AI-generated texts, respectively. This is relatively consistent with near zero false positive rates (i.e., roughly 1 in 10,000) reported by the providers of these tools \cite{emi2025false_positives_ai_detectors}.

To further assess the performance of these tools in the wild, we compiled a benchmark dataset from seven web domains spanning the categories of news, lifestyle, entertainment, and technology, that were documented by journalists and experts as AI-generated per a published report on ``AI slop'' in news media \cite{paviour2025ai}. A total of 105 articles were randomly sampled from the reported web domains by selecting fifteen articles from the ``latest news'' section published within the past six months from each of the seven domains. This evaluation step was designed to assess the robustness of Pangram and GPTZero, especially for false negatives. Pangram classified 55.2\% (N=58) of the 105 articles as Highly Likely AI (i.e., AI-generated), with an average confidence score of 1 (i.e., the maximum level of confidence). Of the remaining articles, 16.2\% (N=17) were classified as Mixed (average confidence score = 0.63), and 28.5\% (N=30) were classified as Unlikely AI (average confidence score = 1). In contrast, GPTZero classified 64.7\% (N=68) out of the 105 articles as AI-generated (average confidence score = 91.8\%), 4.7\% (N=5) as Human-AI mix (average confidence score = 93\%), and 30.4\% (N=32) as Human (average confidence score=92\%). 

However, because both Pangram and GPTZero considers text identified as Mixed and Human-AI Mixed, respectively, to involve the use of AI, we combined the scores of those instances with the scores of instances classified by Pangram and GPTZero as AI-generated for the same input text, and if the combined score exceeds the confidence score assigned for the respective human category for each tool for that input, then we re-classify each of those instances as ``AI'', indicating that these instances are AI-generated (or at least a mix of AI was used), otherwise we re-assign those instances  as ``human''. As a result, 14 out of the 17 instances categorized by Pangram as possibly or likely to involve the use of AI (i.e, Mixed) were re-categorized as AI-generated and only 3 were re-categorized as ``human''. In contrast, all five instances of Human-AI Mixed categorized by GPTZero were re-categorized as ``AI''. Based on the re-categorization of these instances, the new total number of AI-generated instances classified by Pangram and GPTZero is 72 and 73 with an average confidence score of 97.1\% and 91.9\%, respectively. As a result, we observe an 82\% agreement between Pangram and GPTZero in the assigned labels to each text being evaluated in our sample of articles. However, both tools also disagreed on 19 out of 105 articles. GPTZero classified 10 out of 19 articles as AI-generated and 9 as human-generated (average confidence score=72.8\%), whereas Pangram classified 9 articles to be AI-generated and 10 articles to be Unlikely AI (average confidence score = 0.91).

Assuming that the 105 articles are in fact AI-generated (a reasonable assumption based on the prior reporting of the domains producing AI-slop, but one we are not able to further validate) these results imply the false negative rate of Pangram on this benchmark sample is 31.4\%, compared to GPTZero ($\sim$30.4\%), in correctly detecting AI-generated content. While there is still some discrepancy between the two detectors, we decide to take a conservative approach by moving our analysis forward using Pangram given its relatively higher false negative rate, which aligns with the objective of our evaluation looking to identify the lower bound of AI-generated (rather than human curated) source content. While we acknowledge and recognize the implications of relying on a single detector for our findings (see Limitations section), we see the results we present in the next section using this tool as a reasonable estimate of a lower bound.

\begin{table*}[!t]
\small
\centering
\begin{tabular}{lcccc}
\toprule
& \textbf{ChatGPT} & \textbf{Copilot} & \textbf{Perplexity} & \textbf{Gemini} \\
\midrule
\textbf{Avg. Number of Cited Sources Per Response} & 14.68 & 8.44 & 7.92 & 7.26 \\
\textbf{Number of Unique Source Web Domains} & 3,964 & 3,286 & 1,750 & 2,369 \\
\textbf{Number of Unique Cited Sources} & 10,453 & 6,007 & 5,636 & 5,168 \\
\bottomrule
\end{tabular}
\caption{Descriptive statistics of cited sources per response by each generative search engine (GSE) provider.}
\label{tab:descriptive-stats-per-provider}
\end{table*}
\section{Results}

We collected a total of 2,848 responses by submitting 712 queries on politics, health, and the environment to four generative search engines. These responses predominantly include in-text citations (N=2596; 91.2\%), but not always (N=252; 8.8\%). Out of the total number of responses that \textit{do not} have citations (N=252),  11.6\% of all health-related queries (N=120) received a response with no citations, compared to 8.2\% (N=92) of all queries on the environment and 5.7\% (N=40) of all queries on politics having no citations in responses. Although the proportion of responses with no citations is relatively small compared to the total number of collected responses, it begins to shed light on how LLMs powering generative search engines occasionally rely on internal knowledge versus pulling in external sources to help generate responses. 
Despite this infrequent behavior by GSE, relying on an LLM's internal knowledge could expose users to outdated information and increase the risk of exposure to hallucinated and inaccurate information\cite{deng2025cram,luo2025unsafe}. This is particularly critical in high-stakes topics such as health, where users may turn to these systems to find accurate information. In contrast, responses that included in-text citations contained an average of 9 citations per response. Among the audited systems, ChatGPT cited the highest number of unique sources per response, averaging 14.7 sources, followed by Copilot (8.4), Perplexity (7.9), Gemini (7.3), as reported in Table \ref{tab:descriptive-stats-per-provider}.


Reporting citation patterns at the provider level, we find that out of the total number of unique sources in the dataset (N=26,266), ChatGPT with web search accounts for the largest share of cited sources (N=10,453; 39.8\%), followed by Copilot (N=6,007; 22.9\%), Perplexity (N=5,636; 21.5\%), and then Gemini (N=5,168; 19.7\%). At the topic level, we find that responses to environment-related queries include on average 10.9 citations, whereas responses to queries on health and politics, receive 8.4 and 7.7 citations on average, respectively (as described in Table \ref{tab:descriptive-stats-by-topic}). Furthermore, when examining the proportion of cited sources across generative search engines at the topic–provider level, we divide the total number of unique sources found per topic-provider pair by the total number of unique sources for that particular topic from Table \ref{tab:descriptive-stats-by-topic}. We find that ChatGPT accounts for the largest proportion of unique sources cited in responses to environment, health, and politics-related queries (0.46, 0.36, and 0.31, respectively). Also, responses generated by ChatGPT to queries in these topics cite approximately 20, 12, and 10 sources on average per response, respectively, which is more than other providers (see Table \ref{tab:provider-topic-sources} in the Appendix). 

\begin{table}[htbp]
\small
\centering
\setlength{\tabcolsep}{3pt}

\begin{tabular}{p{4.2cm}ccc}
\toprule
& \textbf{Env.} & \textbf{Health} & \textbf{Politics} \\
\midrule
\textbf{Number of Queries} & 280 & 257 & 175 \\
\textbf{Total Unique Domains} & 3,481 & 2,958 & 1,741 \\
\textbf{Total Unique URLs} & 12,252 & 8,640 & 5,375 \\
\textbf{Avg. URLs per Query} & 10.94 & 8.40 & 7.68 \\
\textbf{Proportion of Total Sources} & 0.466 & 0.329 & 0.205 \\
\bottomrule
\end{tabular}

\caption{Descriptive statistics of the unique cited source web domains and sources by topic category across ChatGPT, Copilot, Gemini, and Perplexity.}
\label{tab:descriptive-stats-by-topic}
\end{table}

To measure the concentration of citations across unique source web domains, we use the Gini index \cite{gini1936concentration} which provides a score that ranges from 0 to 1, where 0 indicates perfect equality (i.e., sources are uniformly distributed across all unique source web domains) and 1 indicates perfect inequality (i.e., sources are concentrated around a few source web domains). By calculating the Gini index on the overall distribution of frequencies of the cited source web domains in the data, we find that the citations in aggregate for all four generative search engines are unequally distributed over sources (Gini=0.68), indicating that cited sources show some concentration rather than being uniformly drawn from the pool of sources. Moreover, when evaluating the concentration of source web domains among cited sources per provider, we find ChatGPT to have the highest domain concentration amongst its sources (0.648), followed by Gemini (0.595), Perplexity (0.563), and then Copilot (0.492).


To examine which sources are driving the concentration of citations across and within each generative search engine, we analyze the distribution of citation frequencies at the source web domain level. We find that the 25 most frequently cited source web domains across providers account for 23.8\% of all citations (i.e., head of the distribution), while the remaining 76.2\% are distributed across a highly skewed set of source web domains. Within this long-tail distribution, the majority of source web domains receive very few citations (59.1\% are cited only once and 16.5\% are cited twice) while a small number of source web domains are cited disproportionately often, as illustrated in Figure \ref{fig:citations-side-by-side}(b) in the Appendix (also see Appendix \ref{gini-table} for breakdown of these proportion by provider). This reflects a concentration of many source web domains with very few citations (as illustrated by the upper left peak of Figure \ref{fig:citations-side-by-side}(b)) compared to the cluster of very few source web domains being cited repeatedly, as illustrated in the bottom right corner of Figure \ref{fig:citations-side-by-side}(b). Upon inspecting this trend among sources with very few citations, we see a large proportion of source web domains have very few citations followed by a rapid drop in the proportion of source web domains with 3 or more citations, as illustrated in \ref{fig:citations-side-by-side}(a) in Appendix \ref{fig:fig-concentration}. This observed long-tail in cited source web domains suggests that generative search engines rely on a somewhat narrow set of repeatedly cited domains in the head of the distribution while predominantly surfacing a large number of minimally cited domains.

Specifically, among the top 25 most cited source web domains across providers, Wikipedia is the single most cited domain, accounting for 15.2\% of the total number of citations in the top 25 most cited source web domains. It is worth noting that when aggregating these most cited source web domains by category, government websites collectively account for 33.0\% of citations among the top 25 source web domains, indicating a clear and relatively strong reliance on official sources. In contrast, social media websites, including Reddit and YouTube, account for 8.8\% of citations among the top 25 source web domains. When combined, both government and social media web domains together account for a striking 41.8\% of all citations within the top 25 most frequently cited source web domains. Moreover, we found that academic and scientific sources, including four major journal publishers (e.g., Nature, ScienceDirect, Frontiers, MDPI) and websites (e.g., ResearchGate), together account for 14.6\% of citations. However, we observe minimal representation of news media outlets among the top 25 most cited source web domains. For more details on the top 25 cited source web domains across and within each provider, please refer to Table \ref{t-25} in the Appendix.

\begin{figure}[!t]
    \centering
    \includegraphics[width=\columnwidth]{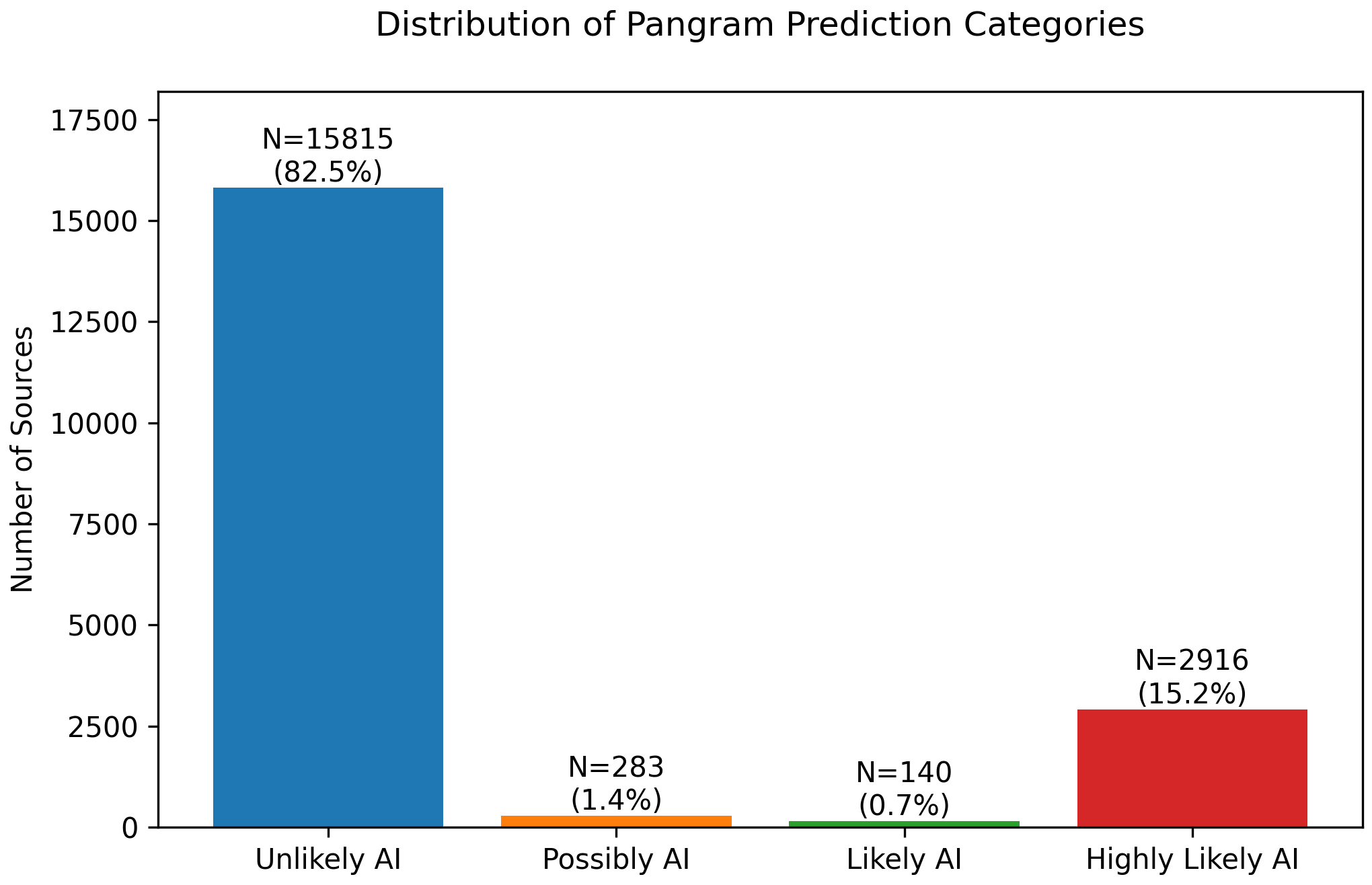}
    \caption{Distribution of Pangram prediction categories.}
    \label{fig:pangram-combined}
\end{figure}

\subsection{Citation of AI-generated Sources}
As previously described in the methods section, we successfully scraped the content of 19,154 unique sources representing 72.9\% of all unique sources in our dataset linking to 6,258 unique source web domains (81.5\% of all unique domains). For each of the 19,154 sources, we submitted the scraped content to Pangram's API to assess whether the text is AI-generated or not and to obtain a prediction score for each prediction category. The content corresponding to the majority of the sources were determined by Pangram to be unlikely generated by AI (i.e.,``Unlikely AI'') (N=15,815; 82.5\%), followed by ``Highly Likely AI'' (N=2916; 15.2\%), ``Possibly AI'' (N=283; 1.4\%), and ``Likely AI'' (N=140; 0.7\%), as illustrated in Figure \ref{fig:pangram-combined}. 

Based on the the distribution of Pangram scores for each prediction category (as shown in Appendix \ref{pangram-scores-dist}), we exclude the ``Possibly AI'' category from further analysis. This conservative decision is based primarily on the difficulty in interpreting this category due to its ambiguity in terms of whether it represents an instance of content that is co-authored with an AI system, or whether it reflects the uncertainty in discerning between human and AI-generated content. In addition, this decision also helps avoid inflating our estimates of AI-generated content in the data given that we have no ground-truth to compare the predictions in this category with\footnote{We sampled 200 URLs classified as Likely AI or Highly Likely AI and manually checked for AI-use disclosure statements, but found, consistent with prior research \cite{russell2025ai}, no instances where such use was disclosed.}. That said, when combining the ``Highly Likley AI'' and ``Likely AI'' categories, we find evidence of a non-negligible share, of approximately 16\% (N=3,056) of cited sources that were successfully scraped, to be AI-generated. Moreover, we found that the top 25 source web domains (i.e., head of the distribution) account for 2.9\% of AI-generated sources as compared to 97.1\% of AI-generated sources that account for the remaining source web domains. This finding further confirms the presence of a very long tail in  source web domains that generative search engines source information from when generating responses to users queries.

To further analyze the data between the various providers of generative search engines, we examined the proportion of unique sources cited by each provider that we were able to scrape and identify as AI-generated, relative to the total number of sources each provider cites across all responses to queries in our sample. We find Copilot to cite AI-generated sources in 27.8\% (N=1,614) of its citations, which translates to nearly 3 out of every 10 cited sources being AI-generated. Similarly, 14.7\% (N=782) of Gemini's cited sources were found to be AI-generated, followed by Perplexity (N=472; 9.4\%), and ChatGPT (N=765; 7.3\%).

By aggregating the data back at the source web domain level, we find that 28.0\% (N=1,754 unique source web domains) out of all unique source web domains (N=6,258) in the sample used for this analysis to have at least one source labeled as AI-generated being cited in generative search engines responses. Looking at the prominence of such content across source web domains with AI-generated sources as identified by Pangram in our sample, we find that 66.2\% of the 1,754 source web domains have one AI-generated source, 17.7\% have 2 AI-generated sources, and all remaining domains (16.0\%) have 3 or more AI-generated sources (as shown in Figure 2 in \ref{dist-ai-gen-cont}). This raises a critical point with respect to a system-design trade-off between offering answers that are substantiated by a diverse set of sources vs. being able to assess the quality of the long tail of sources in terms of how they use AI in the production of content. 

We also examined the top 35 source web domains with the most AI-generated sources cited by generative search engines, and their relative proportions to the total number of cited sources at the source web domain level (see Table \ref{tab:domain-ai-distribution} in the Appendix \ref{ai-gen-dom} for details). This set accounts for 24.7\% of total AI-generated sources in our sample, suggesting that responses for generative search engines tend to be substantiated with AI-generated sources more from the long tail of cited sources. This includes social media sites such as Reddit, government (e.g., pmc.ncbi.nlm.nih.gov, earth.gov) and intergovernmental (e.g., climatepromise.undp.org) sites, as well as a number of sites that are narrowly tailored niche sites (many using .org domain names) about environment and health, and a couple that are news sites (e.g. m.economictimes.com , timesofindia.indiatimes.com). The eighth largest source web domain, with 28 sources detected as AI-generated, was www.factually.co, a website which on its homepage characterizes itself as an ``AI-powered research tool'' and adds that ``Is it perfect? No. Accuracy is hard. Consistency is hard. We're working on it.''

At the topic level, we find AI-generated sources to be the most present in responses to environment-related queries, constituting 13.3\% (N=1,633) of all cited sources in this topic category. This is followed by health-related queries and those on politics where AI-generated sources account for 16.3\% (N=1,408) and 11.1\% (N=592) of all cited sources in each category, respectively.

Switching the analysis to be at topic-provider level, and as reported in Table \ref{num-ai-gen-by-topic-and-provider}, we find Copilot to consistently cite the highest proportion of AI-generated sources across all three topics. In sources cited in responses to environment-related queries, Copilot cites AI-generated sources the most (N=607; 4.9\% of the total citations for this topic category), followed by ChatGPT (N=486; 3.9\%), Gemini (N=365, 2.9\%) and with Perplexity citing the least number of AI-generated sources for this topic category (N=175; 1.4\%). A similar pattern is observed for health-related queries, where Copilot leads all four generative search engines in the number of cited sources with AI-generated sources (N=655; 7.6\% of the total citations for this topic category), followed by Gemini (N=316; 3.6\%), Perplexity (N=237, 2.7\%), and ChatGPT (N=200; 2.3\%) citing the least number of AI-generated sources on this topic. As for cited sources on politics, Copilot was also found to cite the highest number of AI-generated sources (N=352; 6.5\%), followed by Gemini (N=101; 1.8\%), ChatGPT (N=79, 1.4\%), and with Perplexity citing the least number of sources with AI-generated source content (N=60; 1.1\%), as described in Table \ref{num-ai-gen-by-topic-and-provider}. Finally, we report the most frequently cited source web domains per provider for each topic in Table \ref{most-cited-domain-by-provider-within-each-topic} to illustrate the variability between providers with respect to the sources they cite that are AI-generated. For instance, the table shows that Perplexity often leans on Reddit sources that are AI-generated in responding to health-related queries and those on politics, whereas Gemini and ChatGPT reference AI generated sources on health via the nih.gov website. In contrast, Gemini cites AI-generated sources from Times of India (timesofindia.indiatimes.com) in responses related to queries on politics.

\section{Discussion}

This research undertakes an audit of the citations of four popular generative search engines (ChatGPT, Copilot, Gemini, and Perplexity) when responding to users' queries on three topics of public importance: politics, health, and the environment. To do so, we first identified the range of source web domains these engines rely on in responding to users' queries, tabulated their prevalence, identified whether or not the source content cited from those domains is AI-generated, and then we examined the presence of such content among cited sources, topic categories, and providers. Our findings highlight ongoing limitations in the reliability of these engines, especially when responding to queries on critical topics (politics, health, and the environment). For instance the proportion of responses without citations, especially for health-related queries (about 1 in 9), indicates that generative ``search'' engines do rely on their pre-training to answer some questions, raising questions about the timeliness of responses and about instances in which these engines may rely on an LLM's internal knowledge rather than sourcing information from the web. In addition, we reveal insights about the disproportionately concentrated sourcing practices around a handful of source web domains across all four generative search engines. Specifically, we show the degree of concentration of cited sources around a few source web domains, while a preponderance of source citations derived from a long tail of source web domains the majority of which are cited only once or twice. We find evidence that the content of sources linking to some of these domains to be AI-generated, constituting approximately 16\% of all sources cited by generative search engines in response to queries on politics, health, and the environment. In addition, the top 35 source web domains with the highest number of AI-generated source content accounts for 24.7\% of the total AI-generated source content in our sample. This indicates that answer engines substantiate more sources from the long tail less or no clear regards to the quality of such content that users see. 

Our findings raise critical questions about the quality of the information ecosystem as it evolves with the increasing use of generative AI, both by companies using it for generative search responses, as well as by institutions and content-creators using it on the broader internet \cite{kieslichnoyearanticipating-42b, kieslich2024my-4cc}. While AI-generated content is not \textit{necessarily} low quality, known issues with the accuracy of content due to hallucination or gaps in human oversight among professionals \cite{nishal2024envisioning-891}, the potential for large scale influence operations through pollution of the information ecosystem \cite{newsguard_ai_tracking_center}, and emerging demands among end-users for labels to support their agency in opting out of consuming AI-generated content \cite{wittenberg2024labeling}, all suggest that a profusion of AI generated content being referenced in generative search results is a troubling development. Our findings draw further attention to the reliance of generative search engines on social media sources, such as Reddit, and official sources (e.g., government websites), and that additional work is needed to assess the credibility and quality of information on these websites, not only ones which self-disclose their use of AI, but also professional publishers (e.g. Times of India and Economic Times) who may use generative AI in production pipelines with (or perhaps without) appropriate oversight \cite{russell2025ai-620}. Topic specific advocacy sites are proliferating online and finding their way into response answers on environmental and health topics, while the prevalence of government citations raise additional questions about the bias of sources cited \cite{zhang2025source-6e8}. Accordingly, without such assessments focusing on bias, information quality, and credibility of sources,  as well as the contexts in which this type of content is used to help determine its potential risks to users, AI-generated content may continue to circulate on GSEs without sufficient scrutiny. Notably, we found that majority of AI-generated source content from government websites originate from pmc.ncbi.nlm.nih.gov and earth.gov. However, it is unclear whether these instances reflect high-quality or lower-quality AI-generated content that has made its way into official websites.

Overall, this work reflects an initial step towards informing future risk mitigation and source management strategies needed to manage information quality in generative search engines. Our findings can support such efforts to improve the reliability of generative search engines by informing risk assessors and developers about the sourcing practices of these systems through the identification of web domains with AI-generated content, thereby contributing toward enhancing the information quality these systems. While our analysis of AI-generated content is focused on reporting the presence of such content among cited sources, future work focusing on analyzing the characteristics of such sources (e.g., accuracy, writing style, word selection, or claims included in the text) and which of these characteristics is influencing \textit{how} generative search engines embed and surface such content in their final responses would complement our findings by helping to reveal the inner processes of generative search engines related to information selection processes which presently appear to prioritize the number and range of sources over their quality. 

\section{Limitations}

Although our study audits four prominent generative search engines across topics of public importance, it also has its limitations. 
First, our sample consists of queries from the Search Arena dataset \cite{miroyan2025search} that span the domains of politics and health. However, for environment-related queries, this dataset did not include queries that have topical scope beyond weather queries (e.g., ``what is weather like today?'') that fit the definition of this topic, as mentioned earlier in the paper. While this reflects a gap in the Search Arena dataset, which has motivated us to look for Climate Q\&A dataset \cite{de2024climateq} that includes environment-related queries, it does not necessarily indicate that such queries are not present or prominent among users using generative search engines. Rather, it is very plausible that the gap in Search Arena is a result of a bias in the sample, which has implications on the range of topics that are present in (or absent from) users queries. Therefore, we recommend for future studies to acknowledge and try to address biases, beyond topical gaps, in any sample used for auditing generative search engines, including the disproportionate representation of users submitting queries from some regions (e.g., global north) over others (e.g., global south), and the implications of these biases on the characteristics of the sample and interpretability of the results.

Second, we limited queries in our sample to those that are in English language and excluded from them queries that are referring to regions outside of the U.S. All four generative search engines under audit are developed primarily in the United States and the evaluation of cited sources for region-specific queries (e.g., ``What are the current poling numbers for the czech political party ODS?'') or even in other languages is encouraged but would require domain and contextual expertise of the information ecosystem in these regions which the authors do not have. In addition, queries in our sample were limited to single-turn interactions with generative search engines. Although this has facilitated the audit of the four generative search engines through retrieving and analyzing a range of sources that were cited in responses to these queries, future audits should expand on this work to include a deeper analysis of the composition of sources, the quality and bias associated with these sources, and how such sources could vary over time in multi-turn conversational flows as various interaction dynamics may yield different citation patterns by these engines and therefore warrant further analysis.

Third, prior research suggests that the majority of generative search engines, including those in our audit, tend to exhibit left-leaning tendencies when responding to questions on key topics within the categories of politics, health, and the environment \cite{yang2025news,rozado2022_political_orientation_chatgpt,dai2025media}, which limits the generalizability of our findings to systems with potentially different value alignments (e.g., Grok). Accordingly, future audits are encouraged to account for variations across regions, languages, and value systems to understand how such differences may shape citation patterns of sources by generative search engines.

Fourth, although this research is concerned with analyzing text-based content of sources for the presence of AI-generated content, we also note that the remaining 7,112 sources that we could not scrape from the total number of unique sources in our sample (N=26,266) reflect formats and modalities beyond text. For instance, 14.2\% of the sources we could not scrape were URLs linking to PDF files on various websites. Also, we could not fetch the content of 16.2\% of the overall sources cited by generative search engines because they appear to be in other modalities, such as we observed in links to videos\footnote{https://tinyurl.com/3f83mpuv} 
and images\footnote{https://tinyurl.com/mry6f2tm} 
on Facebook. Such instances were not accounted for in our analysis of AI-generated content and therefore our study falls short in analyzing them, though they are surely of interest for future information quality research to assess how information across modalities (some of which may be low quality online memes, marketing, or propaganda) is integrated into responses.

Lastly, our evaluation of AI-detection tools (Pangram and GPTZero) in the wild is based on a curated dataset of AI-generated content from news media that have been identified by journalists. The criteria for selecting which tool to employ in our study was based on assessing the false negative rate for each tool. However, the lack of availability to well established benchmarks that facilitate a more robust evaluation of these tools across various types of content beyond news media (e.g., academic blogs or government websites) and across various domains of use presents another limitation to our evaluation of these tools. Therefore, we invite future research to establish a public leaderboard that reflects the evaluation criteria for these tools and presents their performance across various content types (e.g., news, academic papers) and modalities. Such a resource could help guide researchers in selecting the most performant tool or model depending on their use case and enhance the transparency in this nascent area of research. Moreover, while we attempted to contextualize our evaluation of false positives in Pangram's predictions on our corpus, as articulated in section 4.2, the absence of disclosure statements for the use of AI in content online presents a challenge in truly validating whether content online is AI-generated or is authored by a human using AI to resolve issues with content (e.g., resolving spelling or grammatical mistakes). 
Accordingly, the interpretation of our findings is constrained to the use of Pangram only and that using another AI-detection tool may yield different proportion of AI-generated content among cited sources, especially if these tools vary in their false positive and false negative rates. Thus, we recommend for researchers to present their results from using AI-detection tools as a lower-bound, or potentially ensemble results from various models, especially as these tools continue to evolve over time.

Despite the aforementioned limitations, findings presented in this study highlight a critical and emerging risk: the presence of AI-generated sources among cited sources used by generative search engines, with approximately one in five cited sources on average being AI-generated and used to synthesize responses to user queries. In addition, by disclosing the limitations of this study, we hope future audits of generative search engines can work towards addressing and resolving these limitations in order to have better understanding about the citation dynamics and the proportion of AI-generated content being cited and used on these engines. Lastly, our findings, especially with respect to the presence of AI-generated content on government websites, establish a baseline that future research can compare and contrast their results as they assess the proliferation of AI-generated content online beyond those cited by generative search engines.
\section{Conclusion}

This research audits the citations used by four prominent generative search engines when responding to queries on topics of public importance: politics, health, and the environment. By analyzing the distribution and concentration of cited sources, we find generative search engines to unequally cite sources of information, relying on a narrow set of source web domains some of the time, but more predominantly relying on a long tail of minimally cited source web domains, the majority of which are cited only once or twice. Upon inspecting the content of the cited sources we find approximately 16\% of all unique cited sources to be AI-generated. Through our findings, this work is positioned to serve an initial step towards informing future risk mitigation and source management strategies needed to manage the citation of AI-generated sources in generative search engines. 

As the information ecosystem on the internet continues to evolve, and considering that AI-generated content is not \textit{necessarily} always low in quality, developers of these engines bare partial responsibility in implementing governance strategies to manage information quality of cited sources on their systems. One approach to do so is by enhancing the transparency around the types of sources used, and the quality of their content in generating responses to users' queries. If such mitigation measures are not taken, users of these systems may be exposed to AI-generated content that may contain false or misleading information, potentially resulting in harm particularly in high stakes domains such as health and politics.
\appendix
\setcounter{secnumdepth}{2}
\onecolumn
\section{Appendix}

\subsection{Evaluating AI-detection tools}\label{evals}

To estimate the false positive rate for Pangram and GPTZero, we randomly sampled 200 human written text from the RAID benchmark, the largest and most challenging benchmark dataset for machine-generated text detectors \cite{dugan2024raid}. The RAID dataset includes human-written text from various sources, including abstracts of published research papers, wikipedia pages, reddit posts, and news articles. All human text included in the RAID benchmark were published pre-2022 (i.e., prior to the public release of ChatGPT) and retrieved via the Wayback Machine \cite{wayback_machine}. The 200 human-written texts in our sample were uniformly distributed across the following categories: research abstracts (N=50), wikipedia pages (N=50), reddit posts (N=50), and news articles (N=50). Each human-written text in our sample was sent to Pangram and GPTZero via their respective API to determine whether each content is AI-generated or not. We found that both Pangram and GPTZero have correctly classified all human-written text as ``human'', indicating 100\% accuracy and zero false positive rate.

Next, using the sample of 200 human-written texts, we stratified the dataset by large language model (LLM) and content category (i.e., research abstracts, wikipedia pages, reddit posts, and news articles). We then prompted three models, GPT-5.3 (N=70 samples), Gemini-3 (N=65 samples), and Claude-Haiku-4.5 (N=65 samples),  using the prompt in \ref{prompt-text-generation} to generate text conditioned on the first four words of the leading sentence from each human-written texts. Our selection of the aforementioned three LLMs was based on the fact the AI-generated texts in the RAID benchmark were produced using comparatively older models, whereas our study employs recently deployed frontier models that we leverage for text-generation. Consequently, we suspected that the capabilities of newer models may generate text that is more difficult for AI-detection tools to identify.

Accordingly, the final curated dataset includes 200 AI-generated texts spanning the categories of research abstracts, wikipedia pages, reddit posts, and news articles. We ensured that each LLM generated outputs across all content categories (see Appendix \ref{prompt-text-generation} for the prompt template), following the approach proposed in RAID and Pangram \cite{dugan2024raid, emi2024technical}. In addition, all generated texts were reviewed by the authors to ensure that they did not relate to or explicitly restate the original human-written texts from the RAID benchmark. Each AI-generated text was later sent to Pangram and GPTZero via their respective API for evaluation. Both Pangram and GPTZero detected AI-generated text with 100\% accuracy as being AI-generated.

\subsection{Number of unique sources cited by each Generative Search Engine}\label{tab:provider-topic-sources}

\begin{table}[H]
\centering
\small
\resizebox{\textwidth}{!}{
\begin{tabular}{llcccccc}
\toprule
\shortstack{\textbf{Provider}\\\textbf{}}
& \shortstack{\textbf{Topic}\\\textbf{}}
& \shortstack{\textbf{Number of}\\\textbf{Unique Sources}}
& \shortstack{\textbf{Avg. Sources}\\\textbf{Per Query}}
& \shortstack{\textbf{Prop. of Unique Sources}\\\textbf{(Topic)}}
& \shortstack{\textbf{Prop. of Unique Sources}\\\textbf{(Provider)}}
& \shortstack{\textbf{Prop. of}\\\textbf{All Unique Sources}\\\textbf{(N=26{,}266)}} \\
\midrule
\multirow{3}{*}{ChatGPT}
& Environment   & 5646 & 20.16 & 0.46 & 0.54 & 0.21 \\
& Health    & 3121 & 12.14 & 0.36 & 0.30 & 0.12 \\
& Politics & 1686 & 9.63  & 0.31 & 0.16 & 0.06 \\
\midrule
\multirow{3}{*}{copilot}
& Politics & 1586 & 9.06 & 0.30 & 0.26 & 0.06 \\
& Health    & 2072 & 8.06 & 0.24 & 0.34 & 0.08 \\
& Environment   & 2349 & 8.39 & 0.19 & 0.39 & 0.09 \\
\midrule
\multirow{3}{*}{gemini}
& Environment   & 2714 & 9.69 & 0.22 & 0.53 & 0.10 \\
& Politics & 1179 & 6.74 & 0.22 & 0.23 & 0.04 \\
& Health    & 1744 & 6.79 & 0.20 & 0.34 & 0.07 \\
\midrule
\multirow{3}{*}{perplexity}
& Health    & 1972 & 7.67 & 0.23 & 0.35 & 0.08 \\
& Politics & 1168 & 6.67 & 0.22 & 0.21 & 0.04 \\
& Environment   & 2028 & 7.24 & 0.17 & 0.36 & 0.08 \\
\bottomrule
\end{tabular}
}
\caption{Unique sources cited by each system and topic. To calculate the average number of sources per query, we divide the number of unique sources for each topic by the number of queries corresponding to that topic in the dataset. As described in the Data section, the number of \textit{queries} is 280 for environment, 257 for health, and 175 for politics. In order to calculate the proportion of unique sources per topic, we divide the total number of unique \textit{sources} for each provider-topic pair by the total number of unique sources for each topic (environment is 12,252; health 8640; and politics is 5375).}
\end{table}

\subsection{Top 25 most cited domains across Generative Search Engines}\label{t-25}
\begin{table}[H]
\centering
\begin{tabular}{lclc}
\toprule
\textbf{Domain} 
& \textbf{Number of Citations} 
& \textbf{Category} 
& \textbf{Share of Total Citations(\%)} \\
\midrule
en.wikipedia.org            & 1189 & Encyclopedia                & 3.6354 \\
pmc.ncbi.nlm.nih.gov        & 1078 & Government                  & 3.296 \\
www.sciencedirect.com       & 672  & Research Publisher             & 2.0547 \\
www.reddit.com              & 520  & Social Media                & 1.5899 \\
www.epa.gov                 & 454  & Government                  & 1.3881 \\
www.un.org                  & 357  & International Organization  & 1.0915 \\
www.whitehouse.gov          & 304  & Government                  & 0.9295 \\
www.ipcc.ch                 & 267  & International Organization  & 0.8164 \\
www.climate.gov             & 249  & Government                  & 0.7613 \\
science.nasa.gov            & 228  & Government                  & 0.6971 \\
www.theguardian.com         & 218  & News                        & 0.6665 \\
www.weforum.org             & 215  & International Organization  & 0.6574 \\
www.wri.org                 & 212  & International Organization  & 0.6482 \\
www.nature.com              & 183  & Research Publisher             & 0.5595 \\
www.healthline.com          & 170  & Information Publisher                           & 0.5198 \\
www.youtube.com             & 165  & Social Media                & 0.5045 \\
www.mdpi.com                & 162  & Research Publisher                          & 0.4953 \\
www.britannica.com          & 156  & Encyclopedia                & 0.477 \\
www.pbs.org                 & 150  & News                        & 0.4586 \\
www.frontiersin.org         & 147  & Research Publisher             & 0.4495 \\
www.iea.org                 & 147  & International Organization  & 0.4495 \\
www.researchgate.net        & 141  & Research Platform            & 0.4311 \\
pubmed.ncbi.nlm.nih.gov     & 138  & Government                  & 0.4219 \\
my.clevelandclinic.org      & 132  & Medical Organizations       & 0.4036 \\
www.cdc.gov                 & 131  & Government                  & 0.4005 \\
\bottomrule
\end{tabular}
\caption{Top 25 most frequently cited domains across providers and their corresponding share of total citations in the dataset.}
\label{tab:top-25-domains}
\end{table}

\subsection{Concentration of cited domains by each Generative Search Engine}\label{gini-table}
\begin{table}[H]
\centering
\begin{tabular}{lccc}
\toprule
\textbf{Provider} 
& \textbf{Top 25 Domains' Share of Citations (\%)} 
& \textbf{Gini Index} 
& \textbf{Domains with $\leq$ 2 Citations (\%)} \\
\midrule
ChatGPT      & 29.89\% & 0.648 & 82.81\% \\
Gemini     & 27.92\% & 0.595 & 72.46\% \\
Perplexity & 31.58\% & 0.563 & 88.43\% \\
Copilot    & 15.56\% & 0.493 & 86.93\% \\
\bottomrule
\end{tabular}
\caption{Concentration and inequality of cited domains by provider, ordered by Gini index.}
\label{tab:domain-concentration-gini}
\end{table}

\clearpage
\subsection{Domains with AI-generated content}\label{ai-gen-dom}
\footnotesize
\setlength{\tabcolsep}{1pt} 
\begin{longtable}{p{5cm}ccccccc}
\toprule
\textbf{Domain} 
& \textbf{Highly Likely AI} 
& \textbf{Likely AI} 
& \textbf{Possibly AI} 
& \textbf{Unlikely AI} 
& \textbf{Total URLs} 
& \textbf{AI-Generated URLs} 
& \textbf{Proportion} \\
\midrule
\endfirsthead

\toprule
\textbf{Domain} 
& \textbf{Highly Likely AI} 
& \textbf{Likely AI} 
& \textbf{Possibly AI} 
& \textbf{Unlikely AI} 
& \textbf{Total URLs} 
& \textbf{AI-Generated URLs} 
& \textbf{Proportion} \\
\midrule
\endhead

\highlightrow www.iere.org & 62 & 0 & 0 & 0 & 62 & 62 & 1 \\
www.reddit.com & 61 & 0 & 2 & 457 & 520 & 61 & 0.12 \\
www.pmc.ncbi.nlm.nih.gov & 40 & 4 & 20 & 1014 & 1078 & 44 & 0.041 \\
www.biologyinsights.com & 41 & 0 & 0 & 0 & 41 & 41 & 1 \\
\highlightrow www.climate.sustainability-directory.com & 37 & 0 & 0 & 0 & 37 & 37 & 1 \\
www.timesofindia.indiatimes.com & 30 & 5 & 8 & 35 & 78 & 35 & 0.45 \\
www.greenly.earth & 31 & 0 & 2 & 26 & 59 & 31 & 0.53 \\
\highlightrow www.factually.co & 28 & 0 & 0 & 0 & 28 & 28 & 1 \\
\highlightrow www.numberanalytics.com & 26 & 0 & 0 & 0 & 26 & 26 & 1 \\
\highlightrow www.greenpacks.org & 24 & 0 & 0 & 0 & 24 & 24 & 1 \\
www.earth.org & 24 & 0 & 1 & 103 & 128 & 24 & 0.19 \\
\highlightrow www.maweb.org & 24 & 0 & 0 & 0 & 24 & 24 & 1 \\
\highlightrow www.allaboutamerica.com & 21 & 0 & 0 & 0 & 21 & 21 & 1 \\
www.recovered.org & 19 & 0 & 0 & 1 & 20 & 19 & 0.95 \\
\highlightrow www.pollution.sustainability-directory.com & 17 & 0 & 0 & 0 & 17 & 17 & 1 \\
\highlightrow www.lifestyle.sustainability-directory.com & 16 & 0 & 0 & 0 & 16 & 16 & 1 \\
\highlightrow www.energy.sustainability-directory.com & 16 & 0 & 0 & 0 & 16 & 16 & 1 \\
www.en.wikipedia.org & 14 & 1 & 8 & 1157 & 1180 & 15 & 0.013 \\
\highlightrow www.shunwaste.com & 15 & 0 & 0 & 0 & 15 & 15 & 1 \\
\highlightrow www.theglobalstatistics.com & 15 & 0 & 0 & 0 & 15 & 15 & 1 \\
\highlightrow www.thegoodbug.com & 14 & 1 & 0 & 0 & 15 & 15 & 1 \\
\highlightrow www.climatecosmos.com & 13 & 1 & 0 & 0 & 14 & 14 & 1 \\
www.climateimpact.com & 13 & 0 & 1 & 2 & 16 & 13 & 0.81 \\
\highlightrow www.news.sustainability-directory.com & 13 & 0 & 0 & 0 & 13 & 13 & 1 \\
\highlightrow www.medshun.com & 13 & 0 & 0 & 0 & 13 & 13 & 1 \\
\highlightrow www.sciencenewstoday.org & 13 & 0 & 0 & 0 & 13 & 13 & 1 \\
\highlightrow www.okonrecycling.com & 12 & 0 & 0 & 0 & 12 & 12 & 1 \\
\highlightrow www.globalbioenergy.org & 12 & 0 & 0 & 0 & 12 & 12 & 1 \\
\highlightrow www.solartechonline.com & 12 & 0 & 0 & 0 & 12 & 12 & 1 \\
www.frontiersin.org & 12 & 0 & 3 & 129 & 144 & 12 & 0.083 \\
www.m.economictimes.com & 12 & 0 & 1 & 7 & 20 & 12 & 0.6 \\
www.climatepromise.undp.org & 0 & 12 & 46 & 20 & 78 & 12 & 0.15 \\
www.earth.gov & 11 & 0 & 0 & 66 & 77 & 11 & 0.14 \\
www.finance.yahoo.com & 9 & 2 & 0 & 62 & 73 & 11 & 0.15 \\
\highlightrow www.fiveable.me & 11 & 0 & 0 & 0 & 11 & 11 & 1 \\
\bottomrule
\caption{Domain-level distribution of AI-generated content and their respective proportion out of the total sources cited by generative search engines that were successfully scraped for that domain. The highlighted rows show the domains with 100\% of their cited content to be AI-generated.}
\label{tab:domain-ai-distribution} \\
\end{longtable}
\clearpage
\subsection{Number of unique AI-generated sources by topic and provider}\label{num-ai-gen-by-topic-and-provider}
\begin{table}[H]
\centering
\setlength{\tabcolsep}{8pt}
\begin{tabular}{llcccc}
\toprule
\textbf{Topic Category} 
& \textbf{Provider} 
& \textbf{Unique Sources} 
& \textbf{Total Unique Sources Per Topic} 
& \textbf{Proportion} \\
\midrule
Health     & Copilot      & 655 & 8640  & 7.6 \\
Politics  & Copilot      & 352 & 5375  & 6.5 \\
Environment    & Copilot      & 607 & 12252 & 4.9 \\
Environment    & ChatGPT       & 486 & 12252 & 3.9 \\
Health     & Gemini       & 316 & 8640  & 3.6 \\
Environment    & Gemini       & 365 & 12252 & 2.9 \\
Health     & Perplexity   & 237 & 8640  & 2.7 \\
Health     & ChatGPT       & 200 & 8640  & 2.3 \\
Politics  & Gemini       & 101 & 5375  & 1.8 \\
Politics  & ChatGPT       & 79  & 5375  & 1.4 \\
Environment    & Perplexity   & 175 & 12252 & 1.4  \\
Politics  & Perplexity   & 60  & 5375  & 1.1  \\
\bottomrule
\end{tabular}
\caption{Number of unique sources identified as AI-generated by topic and provider. Values in the table are sorted by the number of unique sources column from from highest to lowest.}
\label{tab:ai-generated-top-domains-by-topic-provider}
\end{table}

\subsection{Most cited domains by Generative Search Engine by topic}\label{most-cited-domain-by-provider-within-each-topic}
\begin{table}[H]
\centering
\begin{tabular}{lllc}
\toprule
\textbf{Topic Category} 
& \textbf{Provider} 
& \textbf{Most Cited Domain} 
& \textbf{Citation Count} \\
\midrule
climate   & Copilot      & www.iere.org                              & 55 \\
climate   & Gemini       & www.climate.sustainability-directory.com  & 15 \\
climate   & ChatGPT       & www.climate.sustainability-directory.com  & 15 \\
climate   & Perplexity   & www.reddit.com                            & 16 \\
health    & Copilot      & www.biologyinsights.com                   & 33 \\
health    & Gemini       & www.pmc.ncbi.nlm.nih.gov                  & 13 \\
health    & ChatGPT       & www.pmc.ncbi.nlm.nih.gov                  & 10 \\
health    & Perplexity   & www.reddit.com                            & 10 \\
political & Copilot      & www.factually.co                          & 21 \\
political & Gemini       & www.timesofindia.indiatimes.com           & 21 \\
political & ChatGPT       & www.en.wikipedia.org                      & 6  \\
political & Perplexity   & www.reddit.com                            & 13 \\
\bottomrule
\end{tabular}
\caption{Most cited domain by provider within each topic.}
\label{tab:top-domain-by-topic-provider}
\end{table}

\clearpage
\subsection{Taxonomy of query intent from Search Arena}\label{q-intent}
\begin{table}[H]
\centering
\begin{tabular}{p{0.22\linewidth} p{0.72\linewidth}}
\toprule
\textbf{Category} & \textbf{Description} \\
\midrule

\textbf{Text Processing} & Users request linguistic tasks such as summarizing, translating, paraphrasing, or refining text. \\[0.5em]

\textbf{Creative Generation} & Users request original creative outputs, including fictional scenarios, satire, poetry, storytelling, or other forms of artistic language generation. \\[0.5em]

\textbf{Factual Lookup} & Users seek to retrieve a precise, objective fact or specific information (what/who/when-like questions). \\[0.5em]

\textbf{Information Synthesis} & Users seek a concise, aggregated summary that combines facts or perspectives from multiple sources. The focus is on integrating factual content without requiring reasoning, subjective interpretation, or personalization. \\[0.5em]

\textbf{Analysis} & Users seek a reasoned judgment or breakdown of a topic, often involving comparison, evaluation, weighing different perspectives, and synthesizing material from many sources. Often open-ended. \\[0.5em]

\textbf{Recommendation} & Users request suggestions or advice tailored to particular constraints, preferences, or specified criteria. Often implies personalization or ranking. \\[0.5em]

\textbf{Explanation} & Users seek detailed clarifications, educational insights, or thorough elaborations aimed at better understanding a concept, process, or phenomenon. May ask ``why'' or ``how'' something works without implying action. \\[0.5em]

\textbf{Guidance} & Users request instructions, procedures, or practical advice intended to accomplish specific tasks, typically involving sequential steps or troubleshooting. Usually framed as “how to” or involving action. \\[0.5em]

\textbf{Other} & Prompts that don't fit neatly into any of the categories above. Prompts may be malformed, declarative in nature, or lacking in meaningful user intent. \\

\bottomrule
\end{tabular}
\caption{Query intent taxonomy and corresponding descriptions based on the Search Arena dataset \cite{miroyan2025search}.}
\label{tab:user_intent_taxonomy}
\end{table}

\subsection{Distribution of Climate Q\&A queries over 15 categories}\label{c-q-n-a}
\begin{table}[ht]
\centering
\renewcommand{\arraystretch}{1.25}
\begin{tabular}{@{}l r r@{}}
\toprule
\textbf{Topics} & \textbf{Number of questions} & \textbf{Share of total} \\
\midrule
Climate Change \& Greenhouse gas emissions & 1477 & 43.1\% \\
Earth system & 416 & 12.1\% \\
Energy & 358 & 10.5\% \\
Citizens \& behavior & 173 & 5.1\% \\
Economics & 171 & 5.0\% \\
IPCC-related questions & 170 & 5.0\% \\
Companies \& industries & 153 & 4.5\% \\
Food \& Agriculture & 104 & 3.0\% \\
Feelings & 101 & 2.9\% \\
Biodiversity & 101 & 2.9\% \\
Politics \& Policies & 70 & 2.0\% \\
Circular economy & 67 & 2.0\% \\
Technologies & 35 & 1.0\% \\
Social issues & 29 & 0.8\% \\
\midrule
Total & 3425 & 100\% \\
\bottomrule
\end{tabular}
\caption{Distribution of Climate Q\&A queries \cite{de2024climateq} over 15 topic categories.}
\label{tab:climate-q-n-a}

\end{table}

\clearpage

\subsection{Distribution of sampled categories from Climate Q\&A dataset}\label{c-q-n-a-sample}
\begin{table}[ht]
\centering
\label{tab:topic_distribution_share}
\begin{tabular}{l r r}
\hline
\textbf{Topic} & \textbf{Count} & \textbf{\% share of total} \\
\hline
Earth system               & 67 & 25.8\% \\
Citizens \& behavior        & 32 & 12.3\% \\
Economics                  & 31 & 11.9\% \\
IPCC-related questions     & 19 & 7.3\% \\
Biodiversity               & 16 & 6.2\% \\
Food \& Agriculture         & 16 & 6.2\% \\
Companies \& industries     & 15 & 5.8\% \\
Circular economy           & 14 & 5.4\% \\
Feelings                   & 13 & 5.0\% \\
Politics \& Policies        & 12 & 4.6\% \\
Technologies               & 10 & 3.8\% \\
Social issues               & 8 & 3.1\% \\
Companies \& Industries      & 7 & 2.7\% \\
\hline
\textbf{Total}              & 260 & 100.0\% \\
\hline
\end{tabular}
\caption{Distribution of sampled categories from the Climate Q\&A dataset \cite{de2024climateq} including the relative percentage for each category out of the total sample size.}
\label{tab:climate-q-n-a-sample}

\end{table}


\clearpage
\subsection{Distribution of AI-generated content by domain}\label{dist-ai-gen-cont}
\begin{figure}[H]
    \centering
    \includegraphics[width=\linewidth]{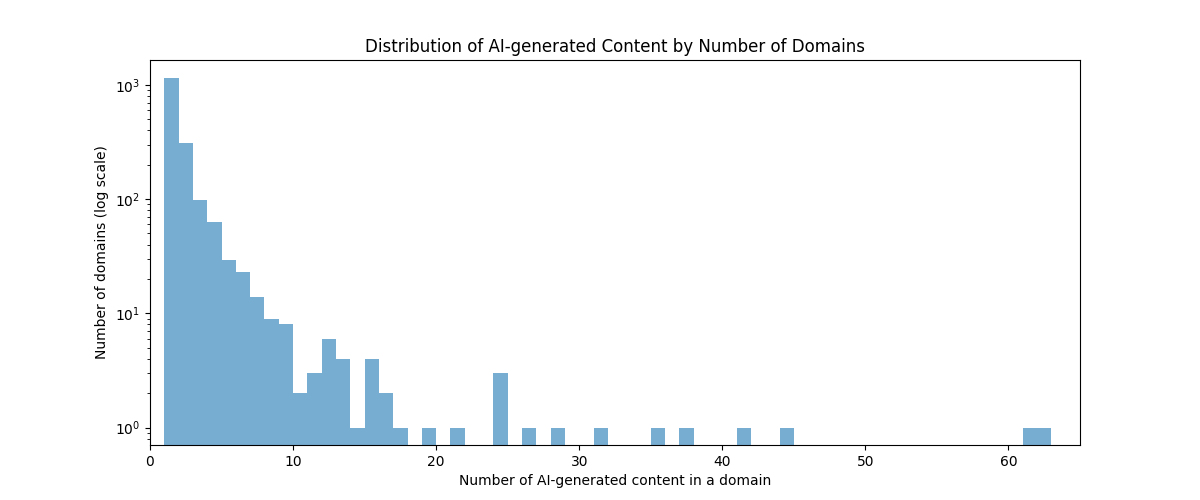}
    \caption{Distribution of AI-generated sources by number of source web domains.}
    \label{fig:ai-generated-content-by-domain}
\end{figure}

\subsection{Prevalence and concentration of cited sources}\label{fig:fig-concentration}
\begin{figure}[H]
    \centering
    \begin{subfigure}[t]{0.48\textwidth}
        \centering
        \includegraphics[width=\linewidth]{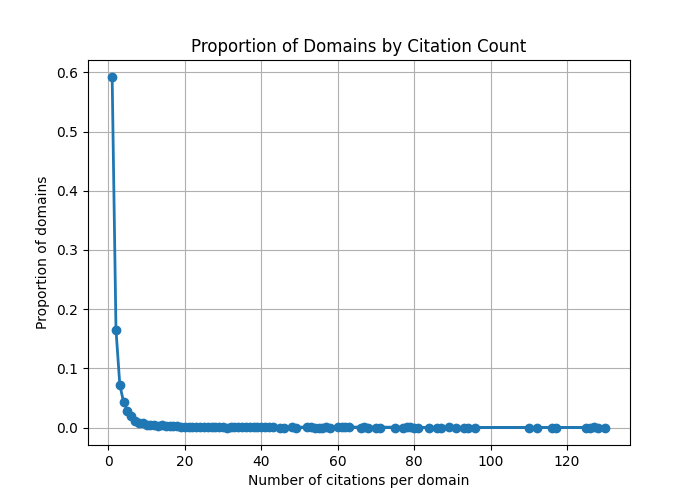}
        \caption{Proportion of Source Web Domains by Citation Count}
        \label{fig:citation-decay}
    \end{subfigure}
    \hfill
    \begin{subfigure}[t]{0.48\textwidth}
        \centering
        \includegraphics[width=\linewidth]{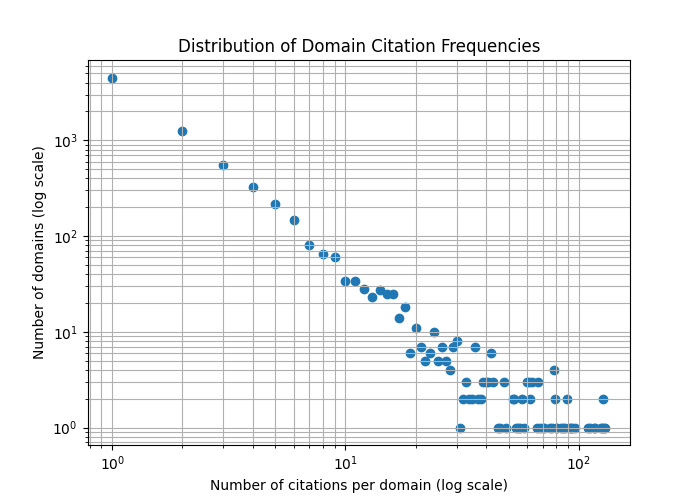}
        \caption{Distribution of Citation Frequencies Across Source Web Domains}
        \label{fig:citation-distribution}
    \end{subfigure}
    \caption{Concentration and distribution of citations across source web domains. Figure 1(a) illustrates the high proportion of source web domains with very few citations and the substantial drop in number of source web domains having 4 or more citations. Figure (b) illustrates the overall distribution of source web domains and citations in our dataset showing a large number of domains with one or two citations, representing the long-tail in our distribution.}
    \label{fig:citations-side-by-side}
\end{figure}
\clearpage
\subsection{Prompt for generating AI-generated text}\label{prompt-text-generation}
\begin{figure}[h]
\centering
\begin{BVerbatim}[frame=single, fontsize=\small]
Write <category> starting with the following four words:
<the first four words of the leading sentence>.
\end{BVerbatim}
\caption{Prompt template used for to generate text across the categories of ``research abstract'', ``wiki page'', ``reddit post'', ``news article''. The four words used in the prompt were retrieved from the leading sentence of the human text in our sample from the RAID benchmark.}
\end{figure}

\subsection{Distribution of Pangram scores}\label{pangram-scores-dist}
\begin{figure}[h]
    \centering
    \includegraphics[width=\textwidth]{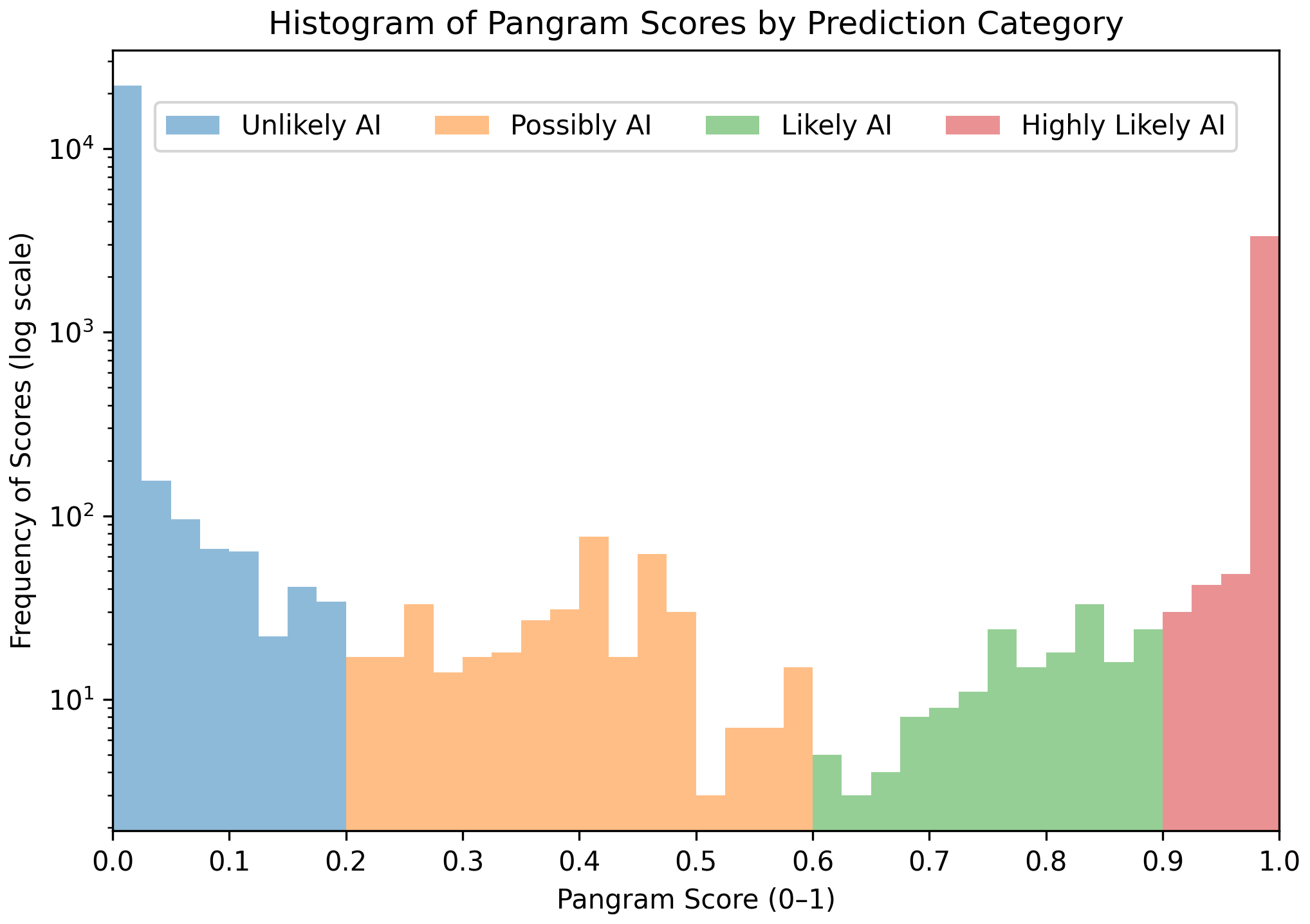}
    \caption{Distribution of Pangram prediction categories.}
    \label{fig:pangram-scores}
\end{figure}

\clearpage
\twocolumn
\bibliography{references}


\end{document}